\documentclass[twocolumn,
nofootinbib,
preprintnumbers,
tightenlines,
superscriptaddress]{revtex4-2}
\usepackage{graphicx} 
\usepackage{mathtools}
\usepackage{amsfonts}
\usepackage{hyperref}
\usepackage{amssymb}
\usepackage{style}
\usepackage{aas_macros}

\newcommand{\ket}[1]{\left| #1 \right\rangle} 

\def\beal{\begin{equation}\begin{aligned}}
\def\eeal{\end{aligned}\end{equation}}

\begin{document}
\title{Boson Cloud Atlas:\\ Direct mass measurements of superradiance clouds near black holes}
\author{Majed Khalaf\,\orcidlink{0000-0001-5537-9992}}
\affiliation{Racah Institute of Physics, Hebrew University of Jerusalem, 91904 Jerusalem, Israel}
\author{Eric Kuflik\,\orcidlink{0000-0003-0455-0467}}
\affiliation{Racah Institute of Physics, Hebrew University of Jerusalem, 91904 Jerusalem, Israel}
\author{Alessandro Lenoci\,\orcidlink{0000-0002-2209-9262}}
\affiliation{Racah Institute of Physics, Hebrew University of Jerusalem, 91904 Jerusalem, Israel}
\author{Nicholas Chamberlain Stone \orcidlink{0000-0002-4337-9458}}
\affiliation{Racah Institute of Physics, Hebrew University of Jerusalem, 91904 Jerusalem, Israel}
\begin{abstract}

Ultralight scalars emerge naturally in several motivated particle physics scenarios and are viable candidates for dark matter.  While laboratory detection of such bosons is challenging, their existence in nature can be imprinted on measurable properties of black holes (BHs).  The phenomenon of superradiance can convert the BH spin kinetic energy into a bound cloud of scalars. In this work, we propose a new technique for directly measuring the mass of a dark cloud around a spinning BH. We compare the measurement of the BH spin obtained with two independent electromagnetic techniques: continuum fitting and iron K$\alpha$ spectroscopy. Since the former technique depends on a dynamical observation of the BH mass while the latter does not, a mismatch between the two measurements can be used to infer the presence of additional extended mass around the BH. We find that a precision of $\sim 1$\% on the two spin measurements is required to exclude the null hypothesis of no dark mass around the BH at a 2$\sigma$ confidence level for dark masses about a few percent of the BH mass, as motivated in some superradiance scenarios. 
\end{abstract}
\maketitle

\section{INTRODUCTION}

Comparisons between two measurements of the same quantity have been powerful tools in the history of astrophysics.  For instance, the first hint of dark matter came from the velocity dispersion of Coma cluster galaxies. Zwicky \cite{Zwicky1933} noted that the dynamical mass of the cluster (obtained from the velocity dispersion) was far larger than the mass estimated from the luminous matter, hinting at the presence of dark matter later evidenced by smaller scale comparisons using rotation curves \cite{Rubin+80}.  
Black holes (BHs) are another type of {non-luminous} mass and their exterior spacetime is thought to be characterized by mass $M$ and spin $J$ only \cite{Carter71}. 
However, many beyond-the-standard-model particle physics scenarios predict extended dark mass around BHs, from density spikes of
dark matter \cite{Young+80, ShapiroPaschalidis14, Kim+23} to ultralight boson fields \cite{Brito+15}. Interestingly, two different techniques to measure BH spin are widely used in modern astrophysics, one of which requires an independent measurement of the mass of the BH (usually obtained from orbital dynamics and Kepler's third law). 
Comparing these spin measurements could reveal the existence of extended dark mass, such as clouds of ultralight bosons.

Ultralight scalars, with masses $\mu \ll$ eV, appear in many extensions of the standard model of particle physics. They are motivated as plausible dark matter candidates \cite{Preskill:1982cy,Abbott:1982af} and may also address several unresolved issues in modern physics, e.g. the strong CP problem \citep{Peccei:1977hh,Kim:1979if,Weinberg:1977ma,Wilczek:1977pj,Shifman:1979if,Zhitnitsky:1980tq,Dine:1981rt} and the hierarchy problem \citep{Arvanitaki+00}. The minimal model where ultralight scalars interact with the standard model solely through the gravitational portal is particularly intriguing \cite{Lenoci:2023gjz}, 
with significant observational implications due to the astrophysical de Broglie wavelength of these scalars (see e.g. ``fuzzy dark matter'' \cite{Hu+00, Goodman00, Hui+17, Mocz+19})

Perturbations to a boson field are unstable in a Kerr background. When the Compton wavelength of the boson field is comparable to the size of the BH horizon, superradiant instability can amplify the boson wave packet  \citep{PressTeukolsky72, Bardeen+72}, forming a \textit{boson cloud}.
The cloud acquires energy and angular momentum at the expense of the central BH, spinning it down.
Once the BH spin decreases sufficiently, the boson cloud stabilizes and reaches its maximum mass, potentially up to $\sim 10$\% of the BH mass $M$ (see e.g. \cite{Baryakhtar:2020gao,Siemonsen:2022yyf}).
The cloud then slowly annihilates via gravitational waves (GW) \cite{Yoshino:2013ofa, Brito+15}.

The efficient transfer of BH spin to the boson cloud can leave a clear imprint on the observed spin demographics of BHs. To date, observed BHs lie in two distinct mass regions: stellar mass BHs ($5 \lesssim M/M_\odot \lesssim 100$; \cite{CasaresJonker14, Abbott+23}) and supermassive BHs (SMBHs; $10^6 \lesssim M/M_\odot \lesssim 10^{10}$; \cite{KormendyHo13}).  
Electromagnetic spin measurements for many observed BHs \citep{Reynolds21} show spins close to extremality, $J\sim GM^2$. This disproves the existence of ultralight scalars with mass $\mu \sim 1/(GM)$, that would dissipate such spin on a timescale given by the instability rate \cite{Arvanitaki:2014wva}. This reasoning applies if the instability timescale is shorter than both the usual timescale for the BH to refill its angular momentum through accretion (the mass-independent Salpeter time $\tau_{\rm S} \sim 50\ {\rm Myr}$) and the age of the system.
More direct observations, such as detecting the continuous GW signal from boson cloud annihilation 
\cite{Arvanitaki:2014wva,PhysRevD.106.042003}, or inference from stellar motions around Sgr~A* \cite{GRAVITY:2023cjt,GRAVITY:2023azi}, have not yet found evidence of such clouds.
 
In this work, for the first time, we show how a subtle difference between the two leading BH spin measurements techniques will in principle enable the direct detection of a boson cloud, if it exists.  We quantify the future improvements in astrophysical spin measurement precision that are required to achieve this goal, finding that $\sim 1\%$ level precision will be necessary for a claim at $2\sigma$. We use natural units $\hbar = c = 1$.

\section{BH Superradiance}\label{sec:SR}

Consider a spinning BH with gravitational radius $R_{\rm g} =GM$ and dimensionless spin $\chi=J/(GM^2)$.
Superradiance (SR) is the exponential amplification of the boson wave packet in the Kerr BH background.
We focus on an ultralight scalar (ULS) with mass $\mu$, working in the non-relativistic approximation. The field then possesses, approximately, a hydrogen-like bound state spectrum, with fine structure constant $\alpha = GM\mu$. SR takes place as long as the ULS wave frequency $\omega$ and the magnetic quantum
number $m\geq 1$ satisfy $
0<\omega\le m \Omega_+$,
where $\Omega_+$ is the angular velocity of the outer BH horizon. When the right-hand inequality is met, we say that the SR condition is saturated. 
This gives $\chi_m(\alpha_m) = [4\alpha_m/m]/[4\alpha_m^2/m^2 + 1]$ (we use subscripts for quantities at saturation for level $m$). Once saturated, energy and angular momentum extraction from the BH ceases.
Away from the BH horizon, the gravitational potential is well approximated by a Newtonian $1/r$ form, reducing the scalar field's motion to a Schr\"{o}dinger equation, whose bound solutions are hydrogen-like wavefunctions $\psi_{n\ell m}({\bf r})$ \cite{Arvanitaki:2010sy,Baumann:2019eav}. A state $\ket{n\ell m}$ forms a \textit{boson cloud} with a probability density peaking at radii $R_c \sim n^2 R_{\rm g}/\alpha^2$ and mass $M_c\equiv\int d^3r\ \mu|\psi_{n\ell m}|^2$.

Due to in-going boundary conditions at the event horizon, the energy spectrum develops an imaginary part: $ \omega_{n\ell m} = E_{n\ell m} + i \Gamma_{n\ell m}$. The real part resembles the spectrum of the hydrogen atom $E_{n\ell m}/\mu = 1-{\alpha^2}/{2n^2} + {\cal O}(\alpha^4)$. This form is consistent with imposing $\chi_m(\alpha_m)\leq 1$, so that the field velocity $v\sim \alpha_m/n\sim \alpha_m/m$, and $\alpha_m/m \le 1/2$. The imaginary part is the instability rate, either decay or SR growth. For small $\alpha$ and $\chi$, the (corrected) Detweiler formula \cite{Detweiler:1980uk} for $\Gamma_{n\ell m}$ is a good approximation \cite{Arvanitaki:2010sy,Pani:2012bp}. However, since we will deal with larger $\chi$ and $\alpha$, we use a next-to-leading order analytical method \cite{Bao:2022hew}, that offers better accuracy ($\lesssim 10\%$ error) when compared to the numerical method from \cite{Dolan:2007mj}. { Note that the mass of the cloud will not depend crucially on relativistic corrections to $\Gamma_{n\ell m}$ as variations of the superradiance time scales will not affect the maximum value of the cloud mass at a given astrophysical timescale (i.e. larger than Myr). At the same time, relativistic corrections to $E_{n\ell m}$ are below $5\%$ for all the values considered in this work \cite{Dolan:2007mj}.}

For a BH with initial mass $M_0$ and spin $\chi_0$, we can compute the final (i) $\alpha$, (ii) $\chi$, and (iii) normalized cloud mass $\zeta \equiv M_c/M$ at the system's age $t=t_{\rm age}$. We relate the final spin $\chi(t_{\rm age})$ to the saturated cloud mass fraction $\zeta(t_{\rm age})$. The reasoning is detailed in the appendix; we consider only the first three SR states ($\ket{211}$, $\ket{322}$, $\ket{433}$) and their rates of GW annihilations. While higher energy states can be considered at the expense of larger $\alpha$, the non-relativistic treatment then becomes invalid. { The calculation of cloud mass directly follows from conservation laws under the assumptions of initial conditions. The only possible uncertainty is the value of the initial BH spin which is not known a priori. In this work, we assume the largest possible value for astrophysical BHs---due to photon capture and radiative angular momentum losses in accretion disks---of $\chi_0=0.998$ \cite{Thorne1974ApJ...191..507T}. This value also gives the best detection prospects. A lower initial value would have a double effect: (i) it would restrict the largest values of $\alpha_0$ we can consider and consequently our access to higher energy superradiant states, and (ii) it would decrease the peak value of the cloud mass, $\zeta$, throughout the system's evolution. However, it will not necessarily have an impact on the value $\zeta(t_{\rm age})$. For example, we checked that when $\chi_0>0.8$, there is no impact on the maximum value of $\zeta(t_{\rm age})$ for $\alpha_0<0.7$. }

\section{Spin measurements}
The no-hair theorem predicts that spin is one of a handful of observables characterizing BH spacetimes \cite{Carter71}.  The two primary methods for measuring BH spins are continuum fitting (CF) \cite{Zhang+97, Shafee+06, McClintock+14} and iron K$\alpha$ line spectroscopy \cite{Fabian+89, Reynolds+14}. Other techniques exist, but are either highly imprecise \cite{Abbott+23}, rely on speculative or incompletely understood models for parameter estimation \cite{StoneLoeb12}, or will only be achievable in the more distant future \cite{AmaroSeoane+23}. Both CF and K$\alpha$ spin measurements assume a  geometrically thin, optically thick disk truncated at the innermost stable circular orbit (ISCO), requirements generally satisfied
for sources accreting with luminosities $(0.01-0.3)$ times the Eddington limit ---common for stellar mass BHs in X-ray binaries (XRBs) and SMBHs in active galactic nuclei (AGN). The CF method further requires that quasi-thermal emission dominates near the ISCO, which occurs in ``soft-state'' XRBs and many tidal disruption events (TDEs) around SMBHs.  

\subsection{X-ray continuum fitting}

The inner regions of many BH accretion disks produce quasi-thermal soft X-ray radiation.  As the temperature increases inwards, and X-rays lie on the Wien tail of the multi-color blackbody spectrum, the X-ray luminosity is highly sensitive to the disk's inner radius, usually assumed to be the ISCO for thin disks. CF applies disk spectrum models to fit observations of continuum radiation from an accreting BH; it ray-traces null geodesics from a general relativistic Novikov-Thorne disk model \cite{NovikovThorne73} through the Kerr spacetime onto the image plane of a distant observer, accounting for weak Comptonization \cite{DavisElAbd19} in the disk atmosphere.  CF therefore measures the physical ISCO radius, $R_{\rm ISCO}$, which can be converted to spin $\chi$ using an independent estimate of the BH mass $M$, i.e. 
$\chi = f(r_{\rm ISCO})$, where $r_{\rm ISCO} \equiv R_{\rm ISCO} / R_{\rm g}(M)$.

Most CF spin estimates come from stellar-mass BHs in XRBs \citep{Shafee+06, Steiner+09, McClintock+14}, where the BH mass estimate is obtained by applying Kepler's 3rd law to the donor star motion \cite{CasaresJonker14}, orbiting at a distance $\sim 10^{6-8}R_{\rm g}$.  As this lies far beyond any plausible boson cloud length scale, the measured BH mass is a {\it dynamical} mass, including not just $M$ but also any extended dark mass around it.  CF is less frequently applied to massive BHs (see below), but when it is used for AGN \citep{Done+13}, the independent BH mass estimate stems from the orbital dynamics of the broad line region \citep{Peterson93, VestergaardPeterson06}, on scales of $\sim 10^4 R_{\rm g}$.  In TDEs around massive BHs, mass estimates stem from gas orbital times at scales of $10^{3-4} R_{\rm g}$ \cite{Rees88, Mockler+19}. So for both AGN and TDEs, the same point about dynamical mass applies, as long as $\alpha$ is sufficiently large
\cite{Arvanitaki:2014wva}.

A basic assumption in CF modeling is that the accretion disk lies in the Kerr equatorial plane.  While this assumption is likely good for long-lived systems like XRBs and AGN, as inclined disks will gradually align themselves into a lower-energy state \cite{BardeenPetterson75}, it may not be true for short-lived systems such as TDEs \cite{StoneLoeb12}. Even for equatorial accretion disks, there are potential theoretical systematics, most notably (i) enhanced Comptonization if magnetic pressure is dynamically important \cite{SalvesenMiller21}, (ii) small deviations from the Novikov-Thorne profile \cite{Kulkarni+11}, and (iii) thermal emission from the plunging, sub-ISCO region \cite{Zhu+12, Mummery+24}. Accurate estimates of system parameters such as distance $D$ and disk inclination $\iota$ are also crucial.

 A practical limitation of CF is that it currently applies primarily to stellar-mass BHs, which reliably exhibit quasi-thermal spectral states. 
 However, there are a minority of AGN for which CF has been attempted \citep{Done+13, Capellupo+17}.
More recently, CF has been used on the highly thermal spectra produced by transient accretion disks in TDEs to measure spin for both SMBHs \citep{Wen+20} and one intermediate-mass BH candidate \citep{Wen+21}. 

\subsection{X-ray reflection spectroscopy}

While many BH accretion disks exhibit the quasi-thermal spectral component suitable for CF, an even greater percentage feature a hard ($\sim 1-10$ keV) X-ray power-law tail originating not from the accretion disk, but from an optically thin corona above it \cite{ZdziarskiGierlinski04}. The non-thermal X-rays irradiate the underlying disk, which absorbs and re-emits a 
reflection spectrum, the most notable feature of which is the Fe K$\alpha$ emission line \cite{Fabian+00}.  The K$\alpha$ line profile is set by a combination of gravitational redshift, Doppler beaming, and strong lensing in the Kerr spacetime; taken together, these relativistic effects encode the spacetime geometry of the K$\alpha$ emission site \cite{Reynolds+14}.  Because most K$\alpha$ flux comes from near the ISCO, K$\alpha$ spectroscopy offers a way to measure the ISCO radius and therefore to infer the spin of the central BH \cite{Iwasawa+96}.  Unlike CF, K$\alpha$ spectroscopy measures the {\it dimensionless} ISCO radius $r_{\rm ISCO}$, 
{\it enabling a spin measurement without an independent mass estimate.}

Aside from the spin, K$\alpha$ spectroscopy must also fit for a number of additional parameters, namely the inclination $\iota$, the ionization parameter $\xi$, and additional phenomenological parameters characterizing the geometry of the coronal irradiation.  
{ X-ray spectroscopy is also affected by a series of  present systematics and uncertainties, that have yet to be addressed unequivocally (see \cite{BAMBI2023} for a recent review). The primary systematics include the uncertainty of the disk geometry, in particular the thickness of the disk \cite{Taylor:2018rlv}, high-density plasma effects \cite{Kallman+21,Ding+24}, uncertainties in the coronal geometry \citep{Chauvin+18} (e.g. between a lamppost geometry and realistic models) that change the illumination pattern see \cite{Wilkins+12},
reflection signatures, thermal \cite{Mummery+24} and reflected \cite{Dong+23}, from inside the plunging region, self-illumination effects \cite{Connors+20} and returning radiation contaminations   \cite{Dauser+22}}

X-ray reflection spectroscopy has been applied across a wide range of BH masses, since  
K$\alpha$ lines are common in both stellar mass XRBs and SMBHs in AGN \cite{Reynolds21}.  Unfortunately, there are no unambiguous detections of K$\alpha$ reflection lines in TDE disk spectra (though see Ref. \cite{Kara+16}); if this is due to the generally weak early time coronal emission in TDEs, future X-ray observations focusing on late time TDEs may identify the K$\alpha$ lines necessary for this technique.

\section{Weighing a boson cloud}

Let us assume we measure a BH spin with the K$\alpha$ method ($\chi_1 \pm \Delta \chi_1$) and with the CF method ($\chi_2 \pm \Delta \chi_2$). Recall that the first technique measures the dimensionless $r_{\rm ISCO}$ while the second obtains the dimensional $R_{\rm ISCO}$. The spin is obtained from $R_{\rm ISCO}$ via an independent dynamical mass measurement $M_{\rm dyn}$. We take as a null hypothesis that $M_{\rm dyn}$ is simply the BH mass $M$. If instead $M_{\rm dyn}= M+M_c$, we can measure the normalized cloud mass
\begin{align} 
\zeta \equiv \frac{M_c}{M} = \frac{g(\chi_1)}{g(\chi_2)}-1,
\end{align}
 where $g$ is the function that enters the Kerr ISCO formula $r_{\rm ISCO}= g(\chi)$. The error on the $\zeta$ estimate can be obtained from error propagation:
\begin{align}
    {\Delta}\zeta &= \sqrt{ \left(\frac{1}{g(\chi_2)}\frac{\partial g}{\partial \chi}\bigg|_{\chi_1}\right)^2\Delta  \chi_1^2 + \left(\frac{g(\chi_1)}{g(\chi_2)^2}\frac{\partial g}{\partial \chi}\bigg|_{\chi_2}\right)^2\Delta  \chi_2^2}\ .
\end{align}

We now assess the observability of $\zeta >0$ situations.  Assuming we know the true spin of the BH, $\chi = \chi_1$ (as K$\alpha$ spin measurements do not suffer from any bias in the dynamical mass), we can estimate the cloud mass $\zeta(t_{\rm age})$. We then compute how the two spin measurements, with errors $\Delta \chi_1 = \Delta \chi_2 = \sigma_\chi$ (symmetry assumed for simplicity), will impact the relative error on $\zeta$, namely $\Delta \zeta/ \zeta$. In this way we can study, in the two-dimensional parameter space $\{ \chi, \sigma_\chi \}$, what are the prospects of observing the cloud with a certain confidence, i.e. to exclude the null hypothesis.

\begin{figure*}
     \centering
\includegraphics[width=.5\textwidth]{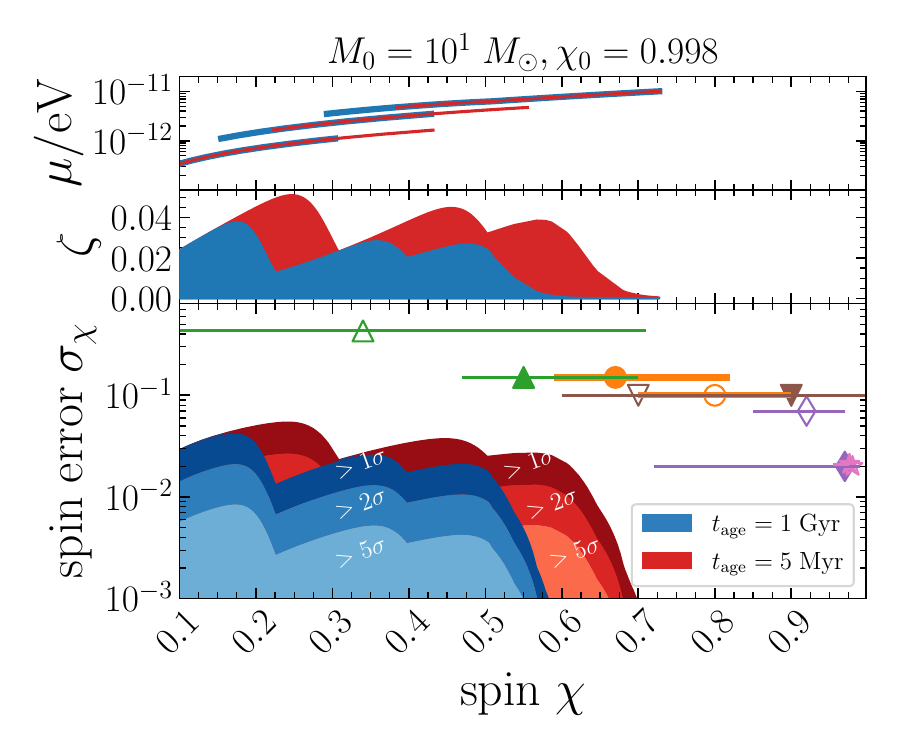} \hspace{-0.03\textwidth}
\includegraphics[width=.5\textwidth]{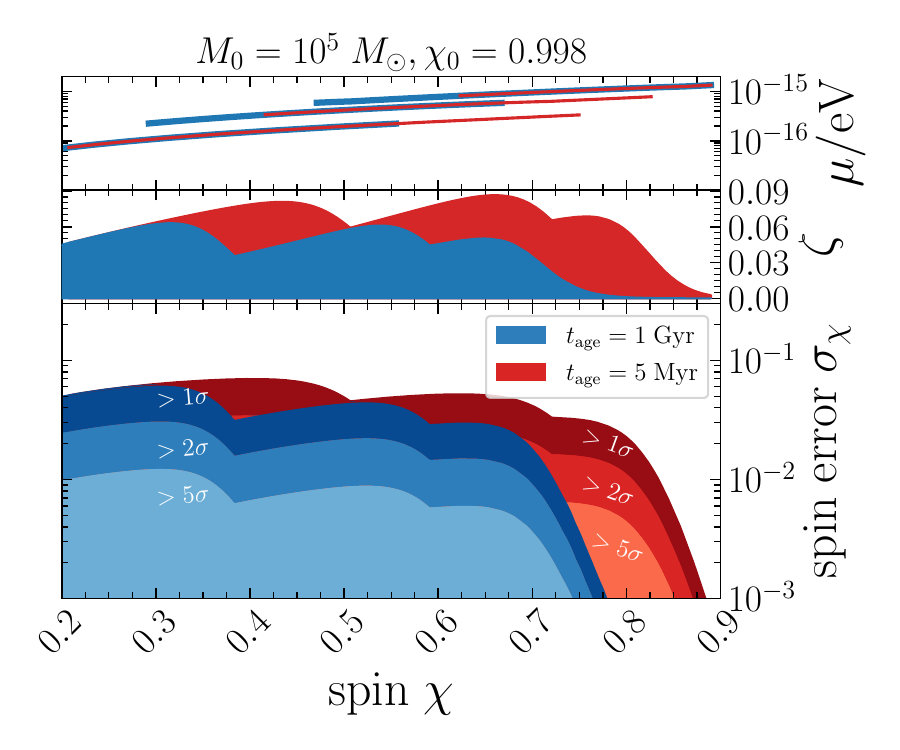} 
     \caption{ Discovery/exclusion potential of our method for stellar mass BHs ({\it left panels}) and massive BHs ({\it right panels}), shown as curves of $\zeta(t_{\rm age})/\Delta\zeta>1,\ 2,\ 5$ in {\it bottom panels}. These curves can be understood as the spin measurement precision $\sigma_\chi$ necessary to achieve 1$\sigma$, 2$\sigma$, and $5\sigma$ detections of a boson cloud with a mass fraction $\zeta$ (shown in {\it middle panels}) produced by an ULS with mass $\mu$ ({\it top panels}).  As the cloud mass evolves over time, we show old systems (1 Gyr, in blue) and young ones (1 Myr, red). The peaks correspond to clouds in $\ket{n\ell m} = \{\ket{211}, \ket{322}, \ket{433}\}$ states, respectively from left to right. The data points represent known XRBs with two independent spin measurements. CF and K$\alpha$ measurements are distinguished by empty and filled markers, respectively. The symbols correspond to 4U 1543-475 (circles), XTE J1550-56 (up triangles), GRO J1655-40 (down triangles), LMC X-1 (diamonds), and GRS1915+105 (stars). }
     \label{fig:cloud_weight}
 \end{figure*}

We show the result in Figure~\ref{fig:cloud_weight}. In the bottom panels, contours show the standard deviations with which the null hypothesis can be excluded in the space of $\{ \chi, \sigma_\chi \}$. We consider both old systems (1 Gyr, e.g. low mass XRBs) and young ones (5 Myr, e.g. high mass XRBs). In the left and right panels we consider the cases of an initial BH of mass 10 $M_\odot$ and $10^5\ M_\odot$, respectively. The sensitivity to $\zeta$ is maximized for the largest cloud mass, with the peaks corresponding to the $\ket{n\ell m} = \{\ket{211}, \ket{322}, \ket{433}\}$ states. We consider a reasonable range of $\mu$ such that $0.01 \leq \alpha_0\leq 0.8$ and find $\chi(t_{\rm age}), \alpha(t_{\rm age}), \zeta(t_{\rm age})$ of the BH using Eqs.~\eqref{eq:cloud_mass_evol}--\eqref{eq:chi_evol} from the appendix. We show in the central panels the final (detectable) normalized cloud mass $\zeta(t_{\rm age})$ and in the upper panels the range of probed boson masses $\mu$, both as functions of $\chi=\chi(t_{\rm age})$. 

For the 10 $M_\odot$ case, we also present existing measurements in the $\{ \chi, \sigma_\chi \}$ plane. As $\sigma_\chi$ is symmetric in our simplified analysis, for the data points we take upper errors for K$\alpha$ measurements and lower errors for CF measurements, with the logic that if an extended dark mass is present, the CF measurement would give a higher spin. 
The take-home message of both plots is that a direct detection of a boson cloud around a stellar mass BH would be possible if the errors on the spins were reduced to $\sigma_\chi \approx 10^{-2}$. For larger BHs, errors $\sigma_\chi \approx {\rm few} \times 10^{-2}$ on a large range of spins may suffice for direct detection, although we do not yet have an individual massive BH system (either TDE or AGN) with both CF and K$\alpha$  spin measurements.  


In Table~\ref{tab:spin_data} we present the stellar mass BHs that, to date, have both measurements of the BH spin, showing the most relevant information for the K$\alpha$ and CF methods. From the quoted measurement errors in this table, we can understand what would be needed in practice to enable a direct detection of a boson cloud: roughly a one order of magnitude improvement in the errors on $M_{\rm dyn}, \ \iota, \ D$.

\begin{table*}[htpb]
\centering
\begin{tabular}{c|cc|ccc||ccccc}
System &Age [Gyr]& Ref&$\chi$ K$\alpha$ & $\iota\ [{\rm deg}]$ & Ref & $\chi$ CF&   $M_{\rm dyn} \ [M_\odot]$ & $\iota\ [{\rm deg}]$ & $D\ [{\rm kpc}]$& Ref  \\\hline
LMC X-1  & 0.005 &\cite{Gou:2011nq}& $0.97^{+0.02}_{-0.25}$ & fixed &\cite{2012MNRAS.427.2552S}& $0.92^{+0.05}_{-0.07}$ & $10.91\pm1.54 $& $36.38 \pm 2.02$ &$ 48.10\pm 2.22$&\cite{2020ApJ...897...84T} \\\hline
4U 1543-475 &$0.1-0.5$&\cite{Fragos:2014cva}& $ 0.67^{+0.15}_{-0.08}$ & $ 36.3^{+5.3}
_{-3.4} $&\cite{2020MNRAS.493.4409D} & $0.8 \pm 0.1$ & $9.4 \pm 0.1$ & $20.7 \pm 1.5$&$7.5 \pm 1.0 $&\cite{2006ApJ...636L.113S}\\
XTE J1550-564 &$4.0-13.5 $&\cite{Fragos:2014cva}& $0.55^{+0.15}_{-0.22}$ & $75 -82$&\cite{2011MNRAS.416..941S} & $0.34^{+0.37}_{-0.43}$ & $9.10 \pm 0.61$ & $74.7 \pm 3. 8$ & $4.38^{+0.58}_{-0.41}$&\cite{2011MNRAS.416..941S}\\
GRO J1655-40 &$0.2-0.6$  &\cite{Fragos:2014cva}&$>0.9$ & $30^{+5}
_{-10}$&\cite{Reis+09} & $0.7 \pm 0.1$ &  $6.30 \pm 0.27$&$70.2 \pm 1.2$&$3.2\pm 0.2$\footnote{However, it has been argued in \cite{2009NewA...14..674F} that $D\lesssim  2$ kpc, driving $\chi_{\rm CF}\sim 0.91$, in agreement with the K$\alpha$ method. }&\cite{2006ApJ...636L.113S}\\
GRS1915+105 & $0.1- 0.9$&\cite{Fragos:2014cva}& $0.976 ^{}_{-0.021 }$ & $67.1^{+1.9}_{-1.1}$ &\cite{2020MNRAS.492..405S} & $>0.98$ & $14.0 \pm 4.4$ & $66\pm 2$ & $11.2 \pm 0.8$&\cite{2006ApJ...652..518M}
\end{tabular}
\caption{List of the systems for which the BH spin has been measured with both K$\alpha$ ($\chi$ K$\alpha$) and CF methods ($\chi$ CF). We include, with references, other important information such as the estimated age of the binary as well as dynamical mass $M_{\rm dyn}$, inclination $\iota$, and distance of the XRB $D$, that are crucial to understand the error budget of the different measurements. Uncertainties are sometimes in 90\% CL and sometimes in 1$\sigma$, we refer the reader to the references for details.}
\label{tab:spin_data}
\end{table*}

\begin{figure}
     \centering
\includegraphics[width=\columnwidth]{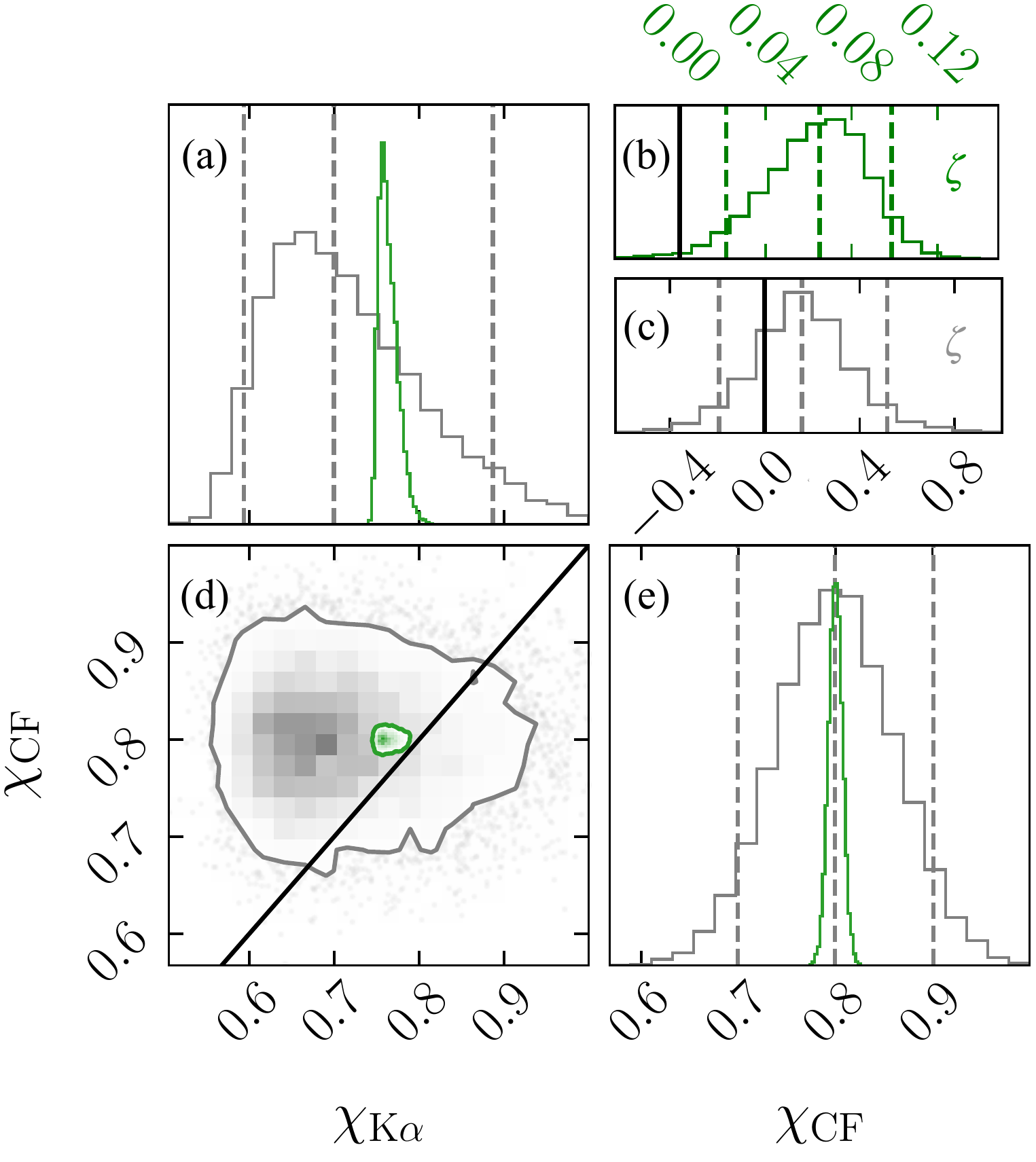}
     \caption{Statistical analysis of mock measurements, inspired by the 4U 1543-475 system. The spin distributions ({\it panels} (a) and (e)) are modelled as skewed Gaussians designed to reproduce the 90\% CL bounds (gray). We superimpose results (green) for distributions with errors reduced by one order of magnitude and  the $\chi_1$ maximum probability value shifted to a value closer to the maximum probability of $\chi_2$. If the black line $\chi_1=\chi_2$ in the probability density contour plot ({\it panel} (d)) does not cross the contours, a detection can be claimed. We infer the likelihood of $\zeta$ ({\it panels} (c) and (b), for the large and small error cases, respectively). }
     \label{fig:likelihood}
 \end{figure}

To show an example of how smaller errors on the spin measurements can allow us to detect a dark extended mass, we perform a statistical analysis inspired by the XRB 4U 1543-475 \citep{2020MNRAS.493.4409D,2006ApJ...636L.113S} Here we will be general and do not consider the dark mass to be necessarlily a boson cloud. We reconstruct toy probability distributions for the spin measurements from K$\alpha$ and CF, enforcing their shape to reproduce actual uncertainties. Then we compute the likelihood of the estimator ${\zeta}$ and we infer our sensitivity to any dark mass: as of now this system is compatible with the null hypothesis, $\zeta=0$. We repeat the analysis with distributions an order of magnitude narrower and artificially shifting the K$\alpha$ measurement closer to the CF one. The result is shown in Figure~\ref{fig:likelihood}. We see that a $>2\sigma$ detection of a dark mass with the smaller errors is possible. The analysis can be repeated in a more rigorous way with actual (posterior) probability distributions for the K$\alpha$ and CF fit results.
 
\section{Discussion and Conclusions}

So far, spin measurements of BHs have {\it indirectly excluded} the existence of scalars with mass $\mu \sim 1/R_{\rm g}$ \citep{Arvanitaki:2014wva,Hoof:2024quk}.
The considered BH mass ranges consistently include some BHs with high spin values, suggesting no bosons with mass $\mu \sim 1/R_{\rm g}$ exist (though this could change with strong self-interactions \cite{Baryakhtar:2020gao}).
In the future, similar arguments might {\it indirectly detect} an ULS by identifying a BH mass range where {\it all} spin measurements are significantly less than unity. While such a scenario would be exciting, it would be hard to conclusively prove the existence of such scalars, as BH spin evolution is highly uncertain due to astrophysical processes \cite{BertiVolonteri08}, and spin-down mechanisms could be argued for certain mass ranges \cite{KingPringle06}.
In contrast, the method we propose offers {\it direct detection} of an extended dark mass from BH superradiance. This approach avoids reliance on uncertain astrophysical growth histories, but requires significant improvements in spin measurement accuracy.

{ The main challenge to implementing our proposed method lies in reducing the statistical and systematic uncertainties in CF and K$\alpha$ BH spin measurements, which currently have errors of at best $\sim 10\%$ \cite{Reynolds21}. While statistical errors could, in principle, be lowered to $\sim1\%$ with more extensive observing campaigns or future X-ray instruments with greater effective area \cite{BarretCappi19}, systematic errors remain more difficult to constrain. Recent theoretical developments have introduced semi-analytic models for X-ray emission from the plunging (sub-ISCO) region \cite{Mummery+24}, which may help to reduce one source of uncertainty. Similarly, advances in radiation magneto-hydrodynamic simulations of accretion flows \cite{Jiang+19, Mishra+22} indicate that relatively simple physical arguments \cite{BegelmanPringle07} could inform estimates of magnetic pressure in accretion disks, offering a path toward more consistent modeling of the spectral hardening factors that dominate CF systematic error budgets \cite{SalvesenMiller21}. In addition, systematic effects such as disk misalignment \cite{Reis+09} might be addressed by introducing additional free parameters into CF  models \cite{Wen+20}, as has been done in K$\alpha$ spectroscopy to treat uncertainties in coronal geometry \cite{Reynolds+14}. Nevertheless, achieving percent-level accuracy will likely require sustained observational and theoretical progress.}

We now conclude by discussing caveats to our work as well as speculations about future directions.

\paragraph*{Larger BH masses.}
For stellar-mass BHs, existing spin measurements already strongly exclude corresponding ULS masses \cite{Arvanitaki:2014wva, Brito+15, Hoof:2024quk}. Our method provides an independent crosscheck of these exclusion regions, though its discovery potential is greater at higher masses, where BH spin distributions are less well understood. 

Massive BH spin measurements are sparser on the Regge plane $\{M,\chi\}$ and mostly rely on applying the K$\alpha$ method to AGN. In recent years, the first attempts to perform CF on AGN have emerged \cite{Done+13}, but in order for these to be useful for direct detection of boson clouds, these will need to (i) rigorously account for spectral distortions from absorption by neutral gas in the host galaxy, and (ii) be applied to AGN with reverberation mapping mass measurements \cite{Peterson93}, as single-epoch broad line mass measurements have far too large of a statistical scatter to be relevant \cite{VestergaardPeterson06}.  
In contrast, TDE disks around SMBHs are nearly ideal for CF \cite{Wen+20}, but they have not yet produced clear K$\alpha$ reflection features.

Intermediate mass BHs 
offer many new orders of magnitude for spin measurements and ULS exclusions (or, optimistically, detections).  
After decades of debate about their existence \citep{Greene+20}, recent evidence now supports these objects as a real BH population \cite{Webb+12, Baldassare+15, Nguyen+19, Wen+21, Haberle+24}. Spin measurements of intermediate mass BHs remain in their infancy \cite{Wen+21}, but the basic considerations are similar to other mass ranges.

\paragraph*{Accretion-powered cloud growth.} In this work, we considered boson clouds growing and annihilating in isolation, neglecting interactions with surrounding accretion disks. However, if an AGN accretes while the cloud is saturated, the cloud's mass could exceed 10\% of the BH mass \cite{Brito+15}. This scenario is intriguing, as larger values of $\zeta$ would make direct detection/exclusion possible with less accurate spin measurements. Due to the increased number of free parameters, we leave a detailed investigation of this scenario for future work.

\paragraph*{General dark matter around BHs.} Finally, we remark that the spin comparison proposed here can probe the presence of any type of extended dark mass, not simply SR clouds. In this work we have neglected self-interactions, but they enable a richer phenomenology, with simultaneous occupation of different SR states \cite{Baryakhtar:2020gao,Collaviti:2024mvh}. Also vector bosons could be interesting as they produce similar boson clouds \cite{Brito+15,Siemonsen:2022yyf}.
Another possibility would be to infer the presence of dark matter spikes, which might emerge naturally in galactic nuclei around massive BHs \cite{Young+80, ShapiroPaschalidis14}, and have even been considered for low mass XRBs \cite{Qin:2024pfu}.

\section*{Acknowledgements}
The authors thank Yifan Chen, Hyungjin Kim, Matthias Koschnitzke, Rotem Ovadia, Ofri Telem, Xiao Xue, Peter Jonker, Assaf Horesh, and Jack Steiner for useful discussions. MK acknowledges the support of the Milner fellowship. MK and AL are grateful to the Azrieli Foundation for the award of an Azrieli Fellowship.
The work of AL and MK is supported by an ERC STG grant (``Light-Dark'', grant No. 101040019). 
This project has received funding from the European Research Council (ERC) under the European Union’s Horizon Europe research and innovation programme (grant agreement Nos. 101040019 and 101125807).  Views and opinions expressed are however those of the author(s) only and do not necessarily reflect those of the European Union. The European Union cannot be held responsible for them. The work of EK, AL and MK is additionally funded in part by the
US-Israeli BSF grant 2020220. EK is additionally supported by NSF-BSF Physics grant 2022713.  NCS acknowledges additional support from the Binational Science Foundation (grant Nos. 2019772 and 2020397) and the Israel Science Foundation (Individual Research Grant Nos. 2565/19 and 2414/23).

\appendix*
\section{Cloud mass}

An important parameter in our analysis is the mass of the cloud produced by superradiance (SR). 
We therefore want a clear prediction for the boson cloud mass depending on the initial conditions of the system, namely the initial black hole (BH) spin $\chi_0$, the initial mass $M_0$ and the boson mass $\mu$. The cloud mass can be written as
\begin{align} 
M_c = \mu \sum_i N_i,
\end{align}
where $N_i$ are the occupation numbers of different states and $i = (n,\ell,m)$. The joint evolution of the BH and cloud system is determined, to a good approximation, by the following differential equations (see e.g. \citep{Baryakhtar:2020gao}):
\begin{align}
    \dot N_i &= 2\Gamma_i N_i - 2\Gamma_{i}^{\rm GW} N_i^2,\\
    \dot M &= - \sum_i 2E_i \Gamma_i N_i + \dot M_{\rm acc},\\
    \dot J &= - \sum_i 2m_i \Gamma_i N_i + \dot J_{\rm acc}.
\end{align}
The $M_{\rm acc}, J_{\rm acc}$ terms define the mass and angular momentum change of the BH due to accretion, respectively \citep{Brito+15}.
We exploit the dimensionless variables $\epsilon_i = N_i/(GM_{0}^2) $, $\alpha = GM\mu$, $\chi = J/(GM^2)$. In this language, the system can be rewritten as
\begin{align}\label{eq:ode_eps}
    \dot \epsilon_i &= 2\Gamma_i \epsilon_i - 2\gamma_{i}^{\rm GW} \epsilon_i^2,\\\label{eq:ode_alpha}
    \dot \alpha &= - \alpha_0^2\sum_i \frac{2E_i}{\mu} \Gamma_i \epsilon_i + \alpha \frac{\dot M_{\rm acc}}{M},\\\label{eq:ode_chi}
    \dot \chi &= -2\chi\frac{\dot \alpha}{\alpha} - \frac{\alpha_0^2}{\alpha^2}\sum_i 2m_i \Gamma_i \epsilon_i + \frac{\dot J_{\rm acc}}{GM^2},
\end{align}
with $\gamma_i^{\rm GW} = \Gamma_i^{\rm GW} GM_{0}^2$. The normalized cloud mass reads
\begin{align}
    \zeta = \frac{M_c}{M} = \frac{\alpha_0^2}{\alpha}\sum_i\epsilon_i\ .
\end{align}

In studying the cloud evolution, we will neglect accretion and in practice consider only a few states, for which SR can happen within the lifetime of the system. Moreover, for fixed $\ell$, since the SR rate of the states with lower $n$ is larger, only states $\ket{n\ell m}\equiv \ket{m+1,m,m}$, with $m\geq 1$, get populated, and only one is occupied at a given time. This holds for the states we are considering in this work, in particular $n<5$.

In the above equation for the occupation number, $\Gamma_i^{\rm GW}$ is the annihilation rate of ultralight scalars into gravitons in the BH background \cite{Yoshino:2013ofa,Arvanitaki:2014wva} $\Gamma_i^{\rm GW}\propto (\mu N_i/M)\mu\alpha^{9+4\ell}$. This process is crucially important for the evolution of the system. If we do not consider gravitational wave (GW) annihilation, the instability happens initially for the most superradiant state allowed by the values $\alpha$ and $\chi$; call it $\ket{m+1,m,m}$. This quickly saturates the cloud, meaning that superradiance will shut off and reach an equilibrium condition, until the superradiant state $\ket{m+2,m+1,m+1}$ is produced, spinning down the BH further and causing the rapid decay of the previous level back into the BH \citep{Arvanitaki:2014wva}. This brings the spin to the initial value and allows for the maximum value of the final cloud mass. In reality, the presence of the GW annihilation term prevents this re-absorption of the lower order state, because the cloud can annihilate into GWs \textit{before} the growth of the next superradiant state even starts.  Therefore the BH spins down from the earlier saturation value $\chi_m$, and forms a lighter cloud than in the case with no annihilations.
 
One can solve numerically the system of equations above for different values of $\alpha_0$ and $\chi_0$ and find the cloud mass at any given moment in time. 
We are particularly interested in the cloud mass when the SR condition is saturated. A simple estimate of this quantity can be obtained as follows.
We assume that $E_i \approx \mu$ (non-relativistic approximation), simplifying the equation for $\dot\alpha$. We neglect the GW annihilation rate, which as we mentioned, has the simple effect of avoiding spin refuelling of the BH. So we can now treat the growth of each cloud state separately, one state at a time and one after the other, from the smallest $m$ allowed by the initial conditions $\alpha_0$ and $\chi_0$ to the largest allowed by the final values after saturation (or by our assumptions on the age of the system). 

\begin{figure*}[th!]
    \centering
    \includegraphics[width=.45\textwidth]{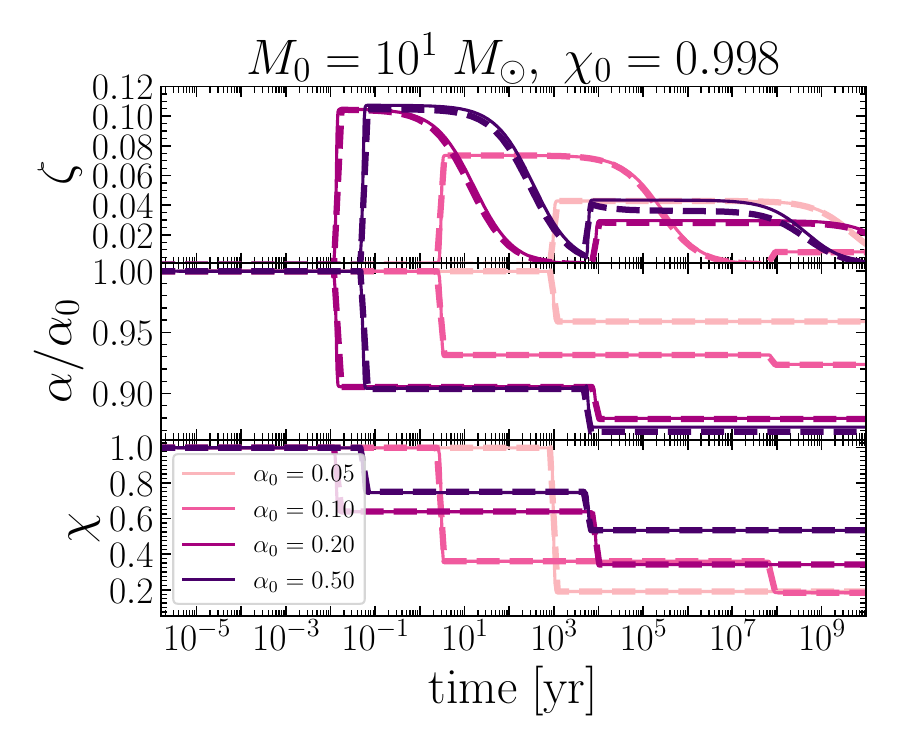}
        \includegraphics[width=.45\textwidth]{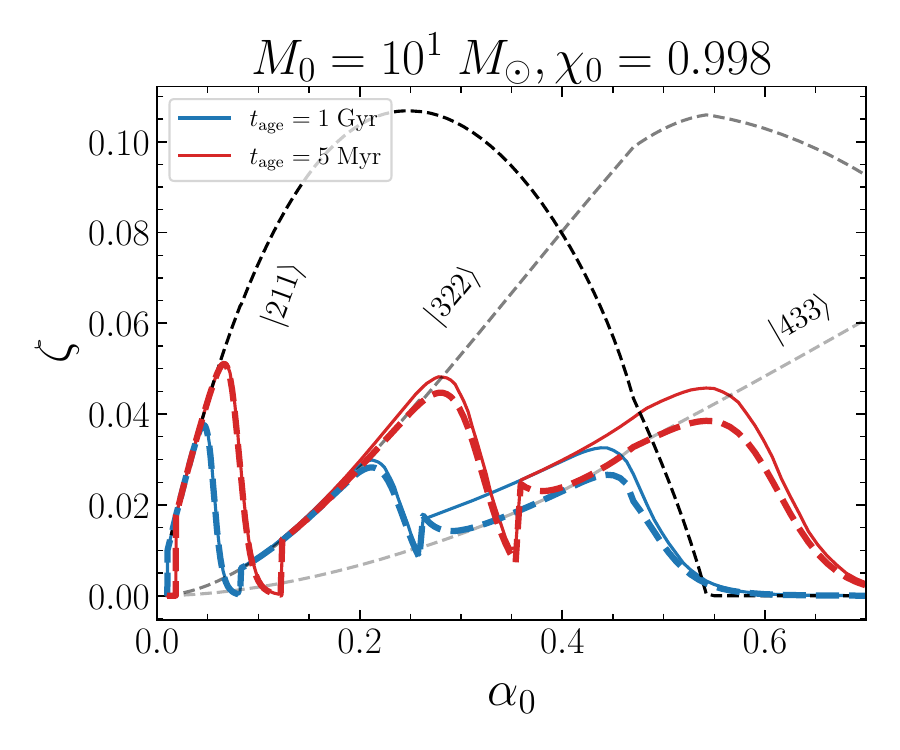}
    \caption{{\it Left panels}: The evolution of the cloud+BH variables $\{\zeta,\ \alpha/\alpha_0,\chi\}$, from {\it top} to {\it bottom} respectively), comparing the numerical result from the system of Eqs.~\eqref{eq:ode_eps}-~\eqref{eq:ode_chi} (solid lines) and the analytical expressions Eqs.~\eqref{eq:cloud_mass_evol}-~\eqref{eq:chi_evol} (thick dashed lines). We choose an initial BH mass of 10 $M_\odot$ and different values of $\alpha_0=\{0.05,0.1,0.2,0.5\}$. The initial spin has been chosen to be $\chi_0=0.998$. {\it Right panel}: The cloud mass as a function of different initial $\alpha_0$ for the same benchmark initial BH. We consider two different ages of the system: 1 Gyr (blue) and 5 Myr (red). We see here how a younger system achieves higher masses. The grey dashed lines show the saturation mass for the different states, solid lines show the numerical result and thick dashed lines are our analytical estimate, used in the reach/exclusion plot and in the main text for the values that satisfy the cloud saturation condition.  }
    \label{fig:cloud_mass}
\end{figure*}

To study the SR of a level $\ket{m+1,m,m}$, we consider a simplified system with no GW annihilation term
\begin{align}
    \dot \alpha &= - \alpha_{m-1}^2     \dot \epsilon_m  ,\\
    \dot \chi &= -\frac{\dot\alpha}{\alpha}\bigg[2\chi - \frac{m}{\alpha}     \bigg] \ .
\end{align}
Imposing $\epsilon_{m-1} = 0$ (we assume that the previous cloud, if any, dissipated before the new one formed), we solve 
\begin{align}\label{eq:alpha_m}
    \alpha_m &= \alpha_{m-1}(1-\alpha_{m-1}\epsilon_m) = \frac{\alpha_{m-1}}{1+\zeta_m}\ ,\\\label{eq:chi_m}
    \chi_m &= \frac{\chi_{m-1}- m\epsilon_m}{(1-\alpha_{m-1}\epsilon_m)^2}\nonumber \\  
    &=  \chi_{m-1}(1+\zeta_m)^2 - \frac{m}{\alpha_{m-1}}  (1+\zeta_m) \zeta_m   \ .
\end{align}
We note that the normalized cloud mass is now $\zeta_m = (\alpha_{m-1}^2/\alpha_m) \epsilon_m$, because $\epsilon_m \equiv N_m /(GM_{m-1}^2)$. To find $\epsilon_m$ we need to use the condition that the cloud is saturated, i.e.  $\chi_m = 4\alpha_m/m/[4\alpha_m^2/m^2+1]$. Equating this result to Eq.~\eqref{eq:chi_m} and using Eq.~\eqref{eq:alpha_m}, one obtains 
\begin{align}\nonumber
   & \zeta_m(\alpha_{m-1},\chi_{m-1})=\\ 
    & -1 + \frac{1+\sqrt{1-16(\alpha_{m-1}/m)^2 (1-\chi_{m-1} \alpha_{m-1}/m)^2}}{2 ( 1-\chi_{m-1} \alpha_{m-1}/m)} \ .
\end{align}
Here it is understood that $\chi_{m-1}> \chi_m$. This is a recursive formula: starting from  $(\alpha_0,\ \chi_0)$ one finds the cloud mass with the minimum $m$ allowed and then uses the second equalities in Eqs.~\eqref{eq:alpha_m}-\eqref{eq:chi_m} to obtain the next values.

Another interesting quantity for us is the cloud mass as a function of the initial and saturated spin (or saturated $\alpha$). To obtain this equation, we use Eqs.~\eqref{eq:alpha_m}-\eqref{eq:chi_m}, solving for $\zeta_m(\chi_m,\chi_{m-1})$ and obtaining
\begin{align}
\zeta_m(\chi_m,\chi_{m-1})=   \frac{\chi_m}{\chi_{m-1}} \frac{1-\sqrt{1-\chi_{m-1} \chi_m}}{1-\sqrt{1-\chi_m^2}} -1\ .
\end{align}
Once the saturation value for the cloud is estimated, we can reintroduce the role of GW annihilation. The cloud mass after saturation of level $\ket{m+1,m,m}$ will evolve as
\begin{equation}
    \zeta_{m,m-1}(t) \equiv \frac{\zeta_m}{1+2\frac{\alpha_m}{\alpha_{m-1}^2}\zeta_m \gamma_m^{\rm GW} t}\ .
\end{equation}

To sum up, the evolution of the system with some initial conditions $(\alpha_0, \ \chi_0)$ is well approximated by the expressions
\begin{align}\label{eq:cloud_mass_evol}
    \zeta(t) &=\Theta(t-\tau_{m_*})    \zeta_{m_*,0}(t-\tau_{m_*}) \nonumber\\
    &\qquad+ \sum_{m=m_*}\Theta(t-\tau_{m})    \zeta_{m,m-1}(t-\tau_m) \ , \\\label{eq:alpha_evol}
    \alpha(t) &=\Theta(t-\tau_{m_*})    \alpha_{0}(t) + \sum_{m=m_*}\Theta(t-\tau_{m})    \alpha_{m}(t) \ , \\
    \label{eq:chi_evol}
    \chi(t) &=\Theta(t-\tau_{m_*})    \chi_{0}(t) + \sum_{m=m_*}\Theta(t-\tau_{m})    \chi{_m}(t) \ .
\end{align}
Here $m_* = \lceil 2\alpha_0\rceil$ is the minimum value of $m$ for which the SR condition can be satisfied given the initial $\alpha_0$ and 
\begin{equation}
    \tau_m \approx \frac{1}{12|\Gamma_m|}\log \bigg[\frac{GM_0^2(\chi_{m}-\chi_{m-1})}{m}\bigg]
\end{equation}
is an estimate of the SR timescale (the factor of 12 is purely empirical). The evolution goes on until the SR condition cannot be satisfied, or until $\tau_m \gtrsim t_{\rm age}$, i.e. the SR timescale becomes larger than the age of the system.

The comparison of the full numerical solution and these analytical estimates is illustrated in Figure~\ref{fig:cloud_mass} for different values of initial fine structure constant $\alpha_0$. We see that the analytical method approximates well the SR growth timescale, the saturation values, the GW annihilation of the cloud and the subsequent SR growth. Small discrepancies are due to the fact that in the numerical solution, the full $E_{n\ell m}$ is considered, while the analytical estimate is done using $E_{n\ell m} \approx \mu$ for simplicity. This is particularly evident for the largest $\alpha_0$ considered. However, in all the examined cases, our approach appears to be reliable and conservative in estimating the cloud mass.

In the right panel of Figure~\ref{fig:cloud_mass} we show the normalized cloud mass as a function of the initial value of $\alpha$ extracted at $t=t_{\rm age} = \{1\ {\rm Gyr}, 5\ {\rm Myr}\}$. We observe three peaks corresponding to the states ($\ket{211}$, $\ket{322}$, $\ket{433}$), and we compare these with the maximum value of the cloud mass (dashed gray lines). The dashed thick lines, representing the analytical approximation, match well with the numerical results and provide a conservative estimate of the cloud mass at the largest values of $\alpha_0$.

 \bibliographystyle{utphys}
\bibliography{ref}

\end{document}